\documentclass[amssymb,aps,twocolumn,showpacs,nofootinbib,superscriptaddress]{revtex4}
\usepackage[]{graphicx}
\usepackage[]{amsmath}
\usepackage{hyperref}

\def\ket#1{| #1 \rangle}
\def\bra#1{\langle #1 |}

\def\kb#1#2{| #1 \rangle\!\langle #2 |}

\def\cE{\mathcal{E}}

\def\uu{{\cal U}}

\def\Tr{\mathrm{Tr}}

\newcommand{\be}{\begin{eqnarray}}
\newcommand{\ee}{\end{eqnarray}}

\def\demi{\frac{1}{2} }
\def\dag{^\dagger }
\begin{document}

\title{Quantum reference frames and deformed symmetries}
\author{Florian Girelli}
\email{girelli@sissa.it}
\affiliation{SISSA, Via Beirut 2-4, 34014 Trieste, Italy and INFN, Sezione di Trieste}
\author{David Poulin}
\email{dpoulin@ist.caltech.edu}
\affiliation{Center for the Physics of Information, California Institute of Technology, Pasadena CA 91125, USA}

\date{\today}

\begin{abstract}
In the context of constrained quantum mechanics,  reference systems are used to construct relational observables that are invariant under the action of the symmetry group. Upon measurement of a relational observable, the reference system undergoes an unavoidable measurement ``back-action" that modifies its properties. In a quantum-gravitational setting, it has been argued that such a back-action may produce effects that are described at an effective level as a form of deformed (or doubly) special relativity. We examine this possibility using a simple constrained system that has been extensively studied in the context of quantum information.  While our conclusions support the idea of a symmetry deformation, they also reveal a host of other effects that may be relevant to the context of quantum gravity, and could potentially conceal the symmetry deformation.
\end{abstract}


\maketitle

\section{Introduction}
\label{sec:Intro}

A common problem faced by all approaches to quantum gravity (QG) is the derivation of a semi-classical theory. There are two aspects to this problem. On the one hand, the theory must predict a flat space-time in a well defined limiting regime. On the other hand, the leading corrections of the theory in this flat space-time approximation need to be well understood and characterized. These corrections are crucial as they establishes a connection to observations, either e.g. in astrophysical phenomenon \cite{Ame02a} or in particle accelerators \cite{HKB+05a}.

The problem of deriving a low energy limit is not specific to QG but rather common in many other areas of physics. In many cases, constructing an {\it ad hoc} effective low-energy theory rather than deriving it from a more  fundamental one has been quite successful. In condensed matter physics for instance, phenomenological theories can very accurately describe exotic phases of matter, e.g. \cite{ZHK89a}, which are otherwise difficult to understand from a more fundamental standpoint. What distinguishes QG is of course the rarity of empirical data. Nonetheless, general principles and physical intuition can guide the construction of such effective theories, and their physical prediction can be confronted with more fundamental QG theories and available experimental data.

An area of research that goes along these lines is deformed special relativity (DSR)\cite{Ame02b,Ame02c,Kow04a,Kow06a}. Since flat space-time is characterized by its symmetries --- namely the Poincar\'e group --- one can modify these symmetries in order to describe potential QG corrections. This approach has the advantage of preserving the relativity principle, albeit through a non-linear realization of the symmetries. The consequences of DSR are not yet fully understood and the theory still undergoes some important developments.

Recently, it has been suggested that DSR may be derived as an effective theory of measurement in a fluctuating quantum space-time \cite{LSV05a,AGG+05a,AGG+06a}. The key insight is that observables in QG are relational --- constructed using physical reference frames (RF). The kinematical coordinates in terms of which the theory is usually formulated are not themselves observable, instead only diffeomorphic invariant relation among them have a direct physical meaning. This aspect of quantum gravity has been discussed at length for many years, see \cite{ROV02a,ROV04a,GMH06a} for overviews. The idea then is that quantum fluctuations of the RF could lead to an effective deformation of the symmetries of observables, which may be expressible as a form of DSR (see also \cite{Smo06a} for a similar argument).

Relational observables are relevant not only to QG, but to any theory endowed with symmetries. Indeed, only the observables that are invariant under the action of the symmetry group --- the {\em constrained} or Dirac observables --- can be probed directly and thus acquire a physical meaning. Quantum relational observables have been discussed quite extensively in the literature \cite{AS67a,AK84a,Rov91a,Mil91a,KMP04a,Pou06a}. Recently, constrained measurements have spurred an interest for quantum RF in the quantum information community, see \cite{PS01a,PS01b,BRS04a,GPP04a,KMP04a,BRST06a,BRS06a,PY07a} for different aspects of the problem.  In this setting, the quantum fluctuations experienced by the RF may limit its use to extract information from other systems, thus placing fundamental limitations on quantum communication and computation. These effects have been studied in great details using exactly solvable models.

In this paper, we build upon this knowledge and examine the possibility of symmetry deformation in a simple constrained model. Our conclusion is that under some circumstances to be specified, the effect of quantum fluctuations on the RF can be interpreted as a deformation of its symmetry group. While the analogy between the model investigated here and QG is far from perfect, we believe that our analysis supports the idea of DSR as an effective low-energy theory. This conclusion is corroborated by some non-trivial properties of many-particle observables common to the two models, which we briefly outline. On the other hand, we also point out a number other consequences of the quantum fluctuations of the RF --- in particular their fundamentally irreversible nature --- that were overlooked in the QG context and may require further investigation.

The remainder of this paper is organized as follows. In the next section, we summarize the proposal \cite{LSV05a,AGG+05a,AGG+06a} of DSR as an effective theory of measurement on a fluctuating quantum  space-time. Section \ref{sec:RRF} presents the constrained model to be investigated in this paper and provides a detailed description of the measurement back-action on the RF \cite{PY07a}.  Section \ref{sec:EXP} addresses the possibility of experimentally detecting the measurement back-action in this simple model, and places fundamental constraints on the regime where it can be detected. The following section addresses the possibility of extending those conclusions to the context of DSR, and also presents other analogies existing between the two models. Finally, Section \ref{sec:conclusion} summarizes our conclusions.

\section{Deformed special relativity}
\label{sec:DSR}

Deformed special relativity is motivated by the existence of a universal quantum gravitational energy scale $M_P$, the Plank mass. Imposing a Planckian maximum value to either  the energy  or the 3-dimensional momentum conflicts with the usual realization of the boost sector of the Lorentz group, which mixes all different length scales. However, a cutoff is consistent with a deformed --- or more precisely non-linear --- realization of the Lorentz symmetry. As a consequence, the mass shell relation are modified to
\begin{equation}
\label{mdr}
E^2=m^2+p^2 \rightarrow E^2= m^2+p^2+ \sum_{n=1}^\infty \alpha_{n}(p,M_ P),
\end{equation}
where $\alpha_{n}$ is a  function of dimension mass squared, and $p=|\vec p|$.

There are various ways to incorporate this deformation, for example using quantum groups technics \cite{MR94a}. A more pedestrian way consists in directly defining a non-linear realization of the Lorentz group on momentum space, leaving the space-time construction aside. To do so, one assumes the existence of an auxiliary momentum variable $\pi$ carrying a linear representation of the Lorentz group. This auxiliary momentum is related the physical momentum $p$ via a non-linear {\it invertible} ``deformation map" :
\be
\label{deformation} p=\uu_{M_P}(\pi). \label{eq:deform_map}
\ee
The choice of $\uu_{M_P}$ is a priori arbitrary, and additional physical consideration and empirical observations are required to constrain it. 
The resulting action of the Lorentz group on $p$ becomes non-linear
\be
\label{boostdeform} p\rightarrow \uu_{M_P} \left(L\cdot \uu_{M_P}^{-1}(p) \right),
\ee
where $L \in SO(3,1)$. The Casimir associated to these deformed symmetries has the general form of Eq.~\eqref{mdr}.

The non-linear relation between $p$ and $\pi$ can always be interpreted as a coordinate change, so {\em a priori} it has no physical consequence. To encode new physics, additional physical inputs are required, and there are many ways to do so. For instance, the action of the Lorentz symmetry on space-time can be specified and, combined with its non-linear action on $p$, it will generally imply a non-trivial addition for the coordinate $p$ for multi-particle states. Non-commutative geometry  \cite{MR94a,Sny47a} is an implementation of that approach. Alternatively, the coordinates $p$ and $\pi$ can be given a direct physical interpretation in terms of other fields. In Refs. \cite{LSV05a,AGG+05a,AGG+06a}, it was suggested that $p$ is the momentum as measured with respect to a physical RF while $\pi$ is the kinematical momentum\footnote{In this setting, the space-time construction is not completely clear, and the authors did not discussed multi-particles states.}. More explicitly, consider a space-time reference frame, that is a tetrad ${e^\mu}_\alpha$, where $\mu$ labels space-time coordinates and $\alpha$  labels different vectors, independent of the chart. If a particle has a momentum $\pi_\mu$, the measured components $p_\alpha$ of this momentum are defined as
\begin{equation}\label{minkows}
p_\alpha= \pi_\mu {e^\mu}_\alpha.
\end{equation}
In Minkowski space-time, the tetrad is trivial ${e^\mu}_\alpha\sim {\delta^\mu}_\alpha$ so $\pi$ and $p$ coincide. Upon change of reference frame\footnote{We consider only Lorentz transformation and leave out translations for sake of clarity.}
\be
 {e^\mu}_\alpha \rightarrow  {\overline e^\mu}_\alpha= {L^\beta}_\alpha {e^\mu}_\beta,
 \ee
the measured components of the momentum are transformed as
 \be
 p'_\alpha= \pi_\mu{\overline e^\mu}_\alpha= {L^\beta}_\alpha {e^\mu}_\beta \pi_\mu= {L^\beta}_\alpha p_\beta .
 \ee
Thus, $p$ carries a linear realization of the Lorentz symmetries. Notice how the Lorentz transformation produces linear combinations among the different vectors labeled by $\alpha$. It is not affecting the space-time indices which are chart dependent and therefore not physical.

To obtain a non-linear realization of the symmetry, we consider a quantum theory and assume that there is an interaction between the RF and the system of interest, leading to a non-trivial ``mixing" between them. Thus, $\pi$, $e_\alpha$, and $p_\alpha$ will henceforth denote quantum operators. Without a complete theory of QG, it is not possible to fully specify what this interaction should be. Nonetheless, it has been argued that an effective treatment of the quantum gravitational fluctuations of the tetrad field can produce such mixing. Aloisio et al. \cite{AGG+06a} modeled a particle traveling in a fluctuating space-time by adding a stochastic term to the tetrad field. The measured momentum, they argued, should then be defined with respect to an average tetrad
\begin{equation}
{E^\mu}_\alpha(e,\pi,M_P) =  \langle{e^\mu}_\alpha\rangle,
\end{equation}
where $\langle \cdot \rangle$ denotes the average of a quantity over a space-time region of size dictated by the de Broglie wavelength of the particle $\lambda = 1/E$.

Making further assumptions about the behavior of the fluctuations, they argued that this average can be encoded in the form of a reversible non-linear map
\be
{e^\mu}_\alpha \xrightarrow{\uu_{M_P}} {E^\mu}_\alpha(e, \pi, M_P).
\label{umap}
\ee
Lorentz transformations act linearly on the tetrad field $e_\alpha$, so their effect on the average tetrad $E_\alpha$ is specified by the following commutative diagram
\be \label{classicaltransf}
\begin{array}{lcl}
 { e^\mu}_\alpha &\stackrel{L}
     {\longrightarrow} &  {\overline e^\mu}_\alpha= {L^\beta}_\alpha {e^\mu}_\beta \\
\downarrow \uu_{M_P}&&\downarrow \uu_{M_P}\\
{ E^\mu}_\alpha &\stackrel{\tilde L}
     {\longrightarrow} & {\overline E^\mu}_\alpha
\end{array}
\ee
This, in turn, induces a non-linear transformation of the measured momentum $p$:
\be
p_\alpha\pi_\mu{ E^\mu}_\alpha \rightarrow p'_\alpha=\pi_\mu{\overline E^\mu}_\alpha =\pi_\mu \uu_{M_P}\left(L \cdot \uu_{M_P}^{-1}
(E)\right).
\label{eq:nlt}
\ee

When the fluctuations have appropriate symmetries, the map $\uu_{M_P}$ takes  the simple form ${E^\mu}_\alpha= F(e,\pi, M_P){e^\mu}_\alpha$, and $F(e,\pi, M_P) \rightarrow 1$ when $M_P\rightarrow\infty$. For example, the Magueijo-Smolin dispersion relation~\cite{MS03a}
\be
\frac{p_0^2-p^2}{1-\frac{p_0}{M_P}}=m^2.
\ee
can be expressed as $p_\alpha = F(e,\pi,M_P) {e^\mu}_\alpha \pi_\mu$ with
\begin{equation}
F(e,\pi, M_P)= \frac{1}{{1-\frac{\pi _\mu {e^\mu}_0}{M_P}}}.
\end{equation}

In short, the physical picture proposed in Refs. \cite{LSV05a,AGG+05a,AGG+06a} is that physical observables are defined with respect to RF, and any interaction between the system of interest and the RF (including interactions required to perform a measurement) will modify  the state of the RF, and hence the value of measured quantities. This picture is quite elegant and largely agrees with the conclusions reached in Ref. \cite{PY07a} in the context of quantum information. Ironically however, the mechanism outlined above does not exactly fit into this picture. The mixing between the system and the RF is caused not by an interaction or measurement, but indirectly by setting a coarse-grained scale that depends on the energy of the measured particle. The model we will consider in the next section adheres more closely to the original philosophy \cite{LSV05a}.

\section{Directional reference frame}
\label{sec:RRF}

\subsection{Quantum gyroscopes}

The previous section presented an argument for the deformation of the Lorentz symmetry caused by a mixing between the system of interest and the RF. To render the analysis of this proposal more tractable, we will consider instead the effect of measurements on a {\em directional} RF, i.e we shift the analysis from $SO(3,1)$ to $SU(2)$. The constraint then is that all physical observables must be rotationally invariant: if $J^\mu$ are the generators of rotation, only observables $O$ with $[O,J^\mu] = 0$ for $\mu = 1,2,3$ are physical\footnote{In appendix A, we provide the  full   algebra of constrained (or Dirac) observables.}. Thus, our primary object of study is basically a gyroscope, a physical system that singles out a particular direction in three-dimensional space.

The quantum analog of a gyroscope is a system with a large amount of spin. The state of the gyroscope is described by a density matrix $\rho$ in the enveloping algebra of the spin-$\ell$ irreducible representation (irrep) of $su(2)$.  Throughout, we will consider $\ell$, or equivalently the Hilbert space dimension $d=2\ell + 1$, as an indication of the \emph{size} of the reference. In practice, gyroscopes are composite systems, built from a large number of elementary particles. For instance the magnetization of a ferromagnet, built up from its component electron spins, can serve as a good gyroscope. In that case, the size $\ell$ of the gyroscope would be proportional to the number of electrons inside the ferromagnetic sample. However, this aspect does not affect our analysis whatsoever, so we will henceforth refer to the gyroscope as if it was a single particle with a large spin.

A gyroscope singles out one direction in three-dimensional space. To obtain a complete directional RF, we need at least two gyroscopes. To conform with the notation of the previous section, we denote the generators of the spin-$\ell$ irrep of $su(2)$ associated to each of these spins by ${e^\mu}_\alpha$  for $\alpha = 1,2$ and $\mu = 1,2,3$. As above,  the index $\alpha$ labels the different gyroscopes (in this case two of them) and $\mu$ is an internal chart dependent index. These operators satisfy the commutation relations
\be
[{e^\mu}_\alpha,{e^\nu}_\beta] =
\left\{\begin{array}{ll}
0 & \rm{if\ } \alpha \neq \beta \\
\frac{i}{\sqrt{\ell(\ell+1)}}{\epsilon_\eta}^{\mu\nu} {e^\eta}_\alpha & \rm{if\ } \alpha=\beta
\end{array}\right.
\label{eq:su2}
\ee
where ${\epsilon_\eta}^{\mu \nu}$ is the totally anti-symmetric tensor with ${\epsilon_1}^{23} = 1$. Note that the unusual factor of $1/\sqrt{\ell(\ell+1)}$ comes in due to the normalization we have chosen. Since each gyroscope is a spin-$\ell$ particle, this normalization is
\be
{e^{\mu}}_\alpha e_{\mu\alpha} = 1, \textrm{ for } \alpha = 1,2.
\ee
The complete  {\em quantum} triad $({e^\mu}_1,{e^\mu}_2,{e^\mu}_3)$ is obtained by defining ${e^\rho}_3 = {\epsilon^\rho}_{\mu\nu} {e^\mu}_1 {e^\nu}_2$. This triad provides a complete quantum RF for three-dimensional space.

The state of each gyroscope $\rho_\alpha$ $\alpha = 1,2$ is a non-negative trace-one operator in the enveloping algebra of the spin-$\ell$ irrep generators ${e^\alpha}_\mu$, i.e. a $d\times d$ matrix. The complete state of the RF is thus $\rho = \rho_1\otimes\rho_2$. Similarly to what was done in previous section, we can define the {\em average triad}
\be
{E^\mu}_\alpha = \langle {e^\mu}_\alpha \rangle = \Tr\{{e^\mu}_\alpha \rho\} .
\label{eq:Esu2}
\ee
Since it will be important later, we point out that this quantum average has a slightly different interpretation than the space-time average used in the previous section. This definition also implies that $\vec E_3 = \vec E_1 \times \vec E_2$. The spatial direction singled out by gyroscope $\alpha$ is thus parallel to $\vec E_\alpha = ({E^1}_\alpha, {E^2}_\alpha, {E^3}_\alpha)$. Note that for $\alpha = 1,2$,
\be
0 \leq \vec E_\alpha \cdot \vec E_\alpha \leq 1-\frac{1}{\ell+1}.
\label{eq:boundE}
\ee
This bound is saturated when the gyroscopes are in so-called {\em coherent states} \cite{Per72a}. In fact, the equality $\vec E^\alpha \cdot \vec E^\alpha = 1-\frac{1}{\ell+1}$ with $E_\alpha$ given by Eq.~\eqref{eq:Esu2} can be taken as the definition of a coherent state $\rho^\alpha$. Coherent states are in some sense most classical as they have the largest amount of spin concentrated in some direction.

\subsection{Quantum measurements}

Gyroscopes can be used to measure the spin of other ``source" particles along the axis of rotation of the gyroscope. We will suppose that these are spin-$j$ particles with $j\ll \ell$. The $2j+1$ dimensional irreducible representation of $su(2)$ is generated by $\pi_\mu$. Once again, the subscript $\mu = 1,2,3$ is a chart-dependent internal index. The state of the source particles are $(2j+1) \times (2j+1)$ density matrix $\sigma$ in the enveloping algebra of the $\pi_\mu$.

Since only rotationally invariant observables have physical meaning, one can used the average triad ${E^\mu}_\alpha$ introduced above to define the {\em semi-classical relational coordinates} of the source particles
\be
p_\alpha = \pi_\mu {E^\mu}_\alpha .
\ee
The semi-classical coordinates can be decomposed into a sum of $2j+1$ orthogonal projectors $P^m_\alpha$, $m=-j,\ldots,j$  associated with distinct eigenvalues $ |\vec E_\alpha |m$. For instance, the semi-classical coordinate associated to the average triad element say $\vec E_\alpha = (0,0,1)$ (which, as a consequence of Eq.~\eqref{eq:boundE}, is only possible when $\ell \rightarrow \infty$) is simply $\pi_3$. A measurement of this semi-classical coordinate has the usual $m=-j,\ldots,j$ outcomes associated to the spin of the particle along the third axis.

One can easily verify the following commutation relation for the semi-classical coordinates
\begin{eqnarray}
[p_1,p_2] &=& ip_3
\label{eq:commp1}\\
\left[p_2,p_3\right] &=& ip_1\vec E_2 \cdot \vec E_2  - i  p_2 \vec E_1 \cdot \vec E_2
\label{eq:commp2}\\
\left[p_3,p_1\right] &=& ip_2\vec E_1 \cdot \vec E_1  - i p_1 \vec E_1 \cdot \vec E_2 .
\label{eq:commp3}
\end{eqnarray}
These relations reduces to the usual $su(2)$ relations when both gyroscopes are in coherent states rotating about perpendicular axes, and $\ell \rightarrow \infty$. (That the two gyroscopes be perpendicular  has no fundamental significance because one can always take different linear combinations of the $p_\alpha$.)

As suggested by the name, the semi-classical coordinates do not take into accound the full quantum nature of the gyroscopes. This is because the triad ${E^\mu}_\alpha$ has the physical meaning of an average, which only acquires an operational meaning in the presence of an ensemble. We thus define the {\em quantum relational coordinates} of the source particles as
\be\label{qcoord}
\mathfrak{p}_\alpha =  \pi_\mu {e^\mu}_\alpha .
\ee
These are physical observables as they commute with the generators of rotation. While all $\pi^\mu$ and $p_\alpha$ are operators on the Hilbert space of the source particle, the operators $\mathfrak{p}_\alpha$ act on the combined Hilbert space of the source particle and the gyroscope $\alpha$. As the different font indicates, the operator $\mathfrak{p}$ does not have a counterpart in the discussion of Sec.~\ref{sec:DSR}. We recover the semi-classical coordinates by taking the average of the quantum coordinates over the state of the gyroscopes
\be
\Tr_{RF}\{\mathfrak{p}_\alpha \rho\} = p_\alpha
\ee
where $\Tr_{RF}$ denotes the partial trace over the Hilbert space of the RF, and as above $\rho = \rho_1\otimes \rho_2$ is the state of the RF. Despite this relation, a measurement of $p_\alpha$ differs from a measurement of $\mathfrak{p}_\alpha$ in two fundamental ways:
\begin{enumerate}
\item The measurement of $\mathfrak{p}_\alpha$ will only be an approximation of what would be obtained by using the corresponding classical reference, i.e. $p_\alpha$.
\item Each time the reference is used to measure $\frak p_\alpha$, it suffers an inevitable ``back-action" which ultimately changes the character of future measurements.
\end{enumerate}
We will describe these points in details, but they can be understood intuitively as follows. The first point is a consequence of the quantum fluctuations of the quantum gyroscope. This effect is minimized when the state of the gyroscope $\rho_\alpha$ is in a coherent state, which minimizes the fluctuations $\sum_\mu \langle {e^\mu}^\alpha \rangle^2$.

The second point is a consequence of the fact that $\mathfrak{p}_\alpha$ does not commute with any of the ${e^\mu}_\alpha$:
\be
[{e^\mu}_\alpha,\mathfrak{p}_\alpha] = \frac i\ell {\epsilon^\mu}_{\nu\eta} {e^\nu}_\alpha \pi^\eta.
\ee
By the uncertainty principle, a measurement of $\mathfrak{p}_\alpha$ will thus alter the value of $e_\alpha$, and so disturb any future measurement that make use of that gyroscope.

The discrepancy between the semi-classical and quantum triads $p_\alpha$ and $\mathfrak{p}_\alpha$ also manifests itself in their commutation relations. The analog of Eqs.~(\ref{eq:commp1}-\ref{eq:commp3}) for the quantum triad is
\begin{eqnarray*}
[\mathfrak{p}_1,\mathfrak{p}_2] &=& i\mathfrak{p}_3 \\
\left[\mathfrak{p}_2,\mathfrak{p}_3\right] &=& i\mathfrak{p}_1 +
-i {e^\mu}_2 e_{\mu 1} \mathfrak{p}_2 +
i\frac{ \frak p_1\frak p_2  -\vec \pi \cdot \vec \pi {e^\mu}_2 e_{\mu 1}}{\sqrt{\ell(\ell+1)} }\\
\left[\mathfrak{p}_3,\mathfrak{p}_1\right]  &=& i\mathfrak{p}_2 +
-i  \mathfrak{p}_1 {e^\mu}_2 e_{\mu 1} +
i\frac{ \frak p_1\frak p_2  -\vec \pi \cdot \vec \pi {e^\mu}_2 e_{\mu 1}}{\sqrt{\ell(\ell+1)} }
\end{eqnarray*}
which differs from Eqs.~(\ref{eq:commp2}-\ref{eq:commp3}) by the addition of a third term, that vanishes in the limit $\ell \rightarrow \infty$.

\subsection{Approximate measurement}
\label{sec:approx}

To understand in what sense a measurement of $\mathfrak{p}_\alpha$ yields an approximation of a measurement of $p_\alpha$ requires basic elements of the theory of generalized measurement. Because of its special nature, the third component of the triad $\frak p_3 = -i[\frak p_1,\frak p_2]$ requires a separate analysis that doesn't provide additional insights into the problem of interest. For simplicity, we will henceforth concentrate on $\alpha = 1,2$.  The spectral theorem can be used to decompose each operator $\mathfrak{p}_\alpha$ as
\be
\frak{p}_\alpha = \sum_{k=\ell-j}^{\ell+j} \lambda^k \Pi_\alpha^k
\ee
where the eigenvalues
\begin{equation}
\lambda^k = \frac{k(k+1)-\ell(\ell+1)-j(j+1)}{2\sqrt{\ell(\ell+1)}}
\label{eq:eig}
\end{equation}
converge to the eigenvalues $k-\ell = -j,\ldots,j$ of $p_\alpha$ when $\ell \rightarrow \infty$. Like the quantum coordinates $\frak{p}_\alpha$ themselves, the spectral projectors  act jointly on the gyroscopes and the source particles. Their explicit form is
\be
\Pi^k_\alpha = \frac{1}{N^k}\prod_{k' \neq k} (\lambda^{k'} - \frak p_\alpha),
\ee
where the normalization factor is
\be
N^k = \prod_{k' \neq k} (\lambda^{k'} - \lambda^k ).
\ee
{\em These projectors are non-linear functions of the relational coordinates} $\frak{p}_\alpha$, more precisely they are polynomials of degree $2j$. When the gyroscopes are in state $\rho$ and the source particle is in state $\sigma$, a measurement of $\frak{p}_\alpha$ will produce the outcomes $\lambda^k$ with probability
\be
Pr(\lambda^k) = \Tr\{\Pi^k_\alpha \rho\otimes\sigma\}.
\ee
This equation can be expressed as
\be
Pr(\lambda^k) = \Tr\{\Lambda^k_\alpha \sigma\}
\label{eq:POVM}
\ee
which only involves the state of the source particles $\sigma$, where the generalized measurement operators $\Lambda^k_\alpha$ are defined as
\be
\Lambda^k_\alpha = \Tr_{RF}\{\Pi^k_\alpha \rho\}.
\ee
Note that $\Lambda^k$ are {\em not} projectors in general, so Eq.~\eqref{eq:POVM} is a generalization of Born's probability rule to the case of positive operator valued measurement (POVM).

When both gyroscopes are in coherent states and $\ell \rightarrow \infty$, the POVM elements $\Lambda^k_\lambda$ are equal to the spectral projectors of the semi-classical observables $P^k_\alpha$ define above. In that limit, measurement of $p_\alpha$ and $\frak{p}_\alpha$ coincide. In general however, the POVM elements $\Lambda^k_\alpha$ are ``coarse-grained" versions of $P^k_\alpha$, mixing the different $k$ components:
\be
\Lambda^k_\alpha &=& (1-\epsilon^k) P_\alpha^k + \sum_{k' \neq k} \epsilon^{k',k} P^{k'}_\alpha
\ee
where all the ``mixing terms" $\epsilon$ are of order $1/\ell$. Thus, any finite quantum RF yields an approximate measurement of the associated semi-classical observable \cite{Wig52a,AY60a}. When the gyroscopes are not in coherent states, the mixing terms become more important. They are of order $1- |\vec E_\alpha |$, c.f. Eq.~\eqref{eq:boundE}.

\subsection{Back-action}\label{sec:backaction}

The second aspect that distinguishes quantum relational coordinates from semi-classical ones is the back-action experienced by the RF. This effect is best described using the generalized theory of quantum dynamics. When a measurement of $\frak p_\alpha$ is performed and outcome $\lambda^k$ is obtained, the joint state the source particle and RF is
\be
\frac{\Pi^k_\alpha (\rho \otimes \sigma) \Pi^k_\alpha}{Pr(\lambda^k)}
\label{eq:proj}
\ee
according to von Neumann's measurement postulate. Averaging over the measurement outcomes and tracing the state of the source particle gives the average state of the RF after the measurement is performed
\be
\rho' = \sum_{k} \Tr_S\{\Pi^k_\alpha (\rho \otimes \sigma) \Pi^k_\alpha\}
\label{eq:CP}
\ee
where $\Tr_S$ denotes the partial trace over the Hilbert space of the source particle. Decomposing the state of the source particle in terms of its eigenstates $\sigma = \sum_i s_i \kb{\phi_i}{\phi_i}$, and performing the partial trace in that same basis gives
\be
\rho' = \sum_{k,i,i'} s_i \bra{\phi_{i'}} \Pi^k_\alpha \ket{\phi_i} \rho \bra{\phi_i} \Pi^k_\alpha \ket{\phi_{i'}}
\ee
Thus, the back action on the RF is described by a dynamical map, which can be written in a Kraus form \cite{Kra83a}
\be
\rho' = \mathcal E_\alpha(\rho) = \sum_a K_a \rho K_a^\dagger
\ee
where for $a = (k,i,i')$ the Kraus operators $K_a = \sqrt{s_j}\bra{\phi_{i'}} \Pi^k_\alpha \ket{\phi_i}$ are operators on the Hilbert space of the RF, and satisfy $\sum_a K_a^\dagger K_a = 1$. The subscript $\alpha$ on $\cE$ reflects the fact that the measurement of different $\frak p_\alpha$ will induce different back-actions on the RF. Expressed in the Heisenberg picture, this map modifies the quantum triad according to
\be
\mathcal E_\alpha^\dagger({e^\mu}_\beta) = \sum_a K_a^\dagger {e^\mu}_\beta K_a.
\ee
By definition of the spectral projectors $\Pi_\alpha^k$ Eq.~\eqref{eq:proj}, {\em the map $\cE_\alpha$ is a non-linear function of the relational coordinates} $\frak p_\alpha$.

Thus we arrive at the conclusion: {\em The measurement of the relational quantum coordinate of a source particle induces a back-action on the RF, which in general is a non-linear function $\cE_\alpha$ of the quantum relational coordinate $\frak p_\alpha$}. This observation supports the proposal of Refs. \cite{LSV05a,AGG+05a,AGG+06a}, but important distinctions will be discussed in Sec.~\ref{sec:discussion}. In what follows, we examine the possibility of detecting this effect experimentally.

\section{Detecting the deformation}
\label{sec:EXP}

A constrained measurement of the spin of a source particle induced a non-linear back-action on the RF.  This will have the effect of changing the measurement outcomes of subsequent measurements that make use of that RF. Since the effect is expected to be tiny, this modification of measurement outcome probabilities can only be detected by repeating the experiment several times and accumulating statistics. Thus, to understand the nature of any experiment aimed at detecting this effect, we must first understand the effect of sequential measurements on the RF.

\subsection{Consecutive relational measurements}

In this section, we describe the dynamics incurred by the RF when it is used to sequentially measure the coordinates of particles drawn from a fixed ensemble. All source particles are assume to be in the same state $\sigma$. Each time the gyroscope $\alpha$ is used to measure the coordinate of a source particle, it experience a back-action described by the quantum map $\cE_\alpha$. After $t$ such measurements, the state of the gyroscope is given by $\cE_\alpha \circ \cE_\alpha \circ \ldots \circ \cE_\alpha(\rho_\alpha) = \cE^t_\alpha(\rho_\alpha)$ where $\circ$ denotes the usual composition.

The effect of this map was studied in great details for the case $j=\frac 12$ in \cite{PY07a}.  In that case, the kinematics is given by the spin-$\frac 12$ generators of $su(2)$, i.e. the Pauli matrices
\begin{equation*}
\pi^1 = \frac 12\left(\begin{array}{cc} 0 &1 \\ 1 & 0 \end{array}\right),\
\pi^2 = \frac 12\left(\begin{array}{cc} 0 &i \\ -i & 0 \end{array}\right),\
\pi^3 = \frac 12\left(\begin{array}{cc} 1 &0 \\ 0 & -1 \end{array}\right).
\end{equation*}
When the gyroscopes' state $\rho_\alpha$ are coherent states and $\ell \gg 1$, the results can be summarized as follows. For $\alpha = 1,2$, the leading order of $\cE_\alpha$ in $1/\ell$ is
\begin{eqnarray}
\cE_\alpha(\rho_\alpha) &\approx& \rho_\alpha -\frac i \ell  \epsilon_{\mu\nu\eta} {E^\mu}_\alpha  \langle \pi^\nu \rangle \sin\theta[{e^\eta}_\alpha,\rho_\alpha] \label{eq:U1}\\
&=& \uu^\dagger_\alpha(e,\pi,1/\ell)(\rho_\alpha) \label{eq:U2}
\end{eqnarray}
where $\cos\theta =  \vec E_\alpha  \cdot \langle \vec \pi \rangle / |\langle \vec \pi \rangle|$. In other words, the RF undergoes a rotation by an angle $\sin\theta |\langle \vec\pi\rangle |/\ell$ about the axis $\vec E_\alpha \times \langle\vec \pi \rangle$ where  $\langle \vec \pi \rangle = \Tr\{\vec \pi \sigma\}$ is the average spin of the source particle. $\uu$ is a unitary rotation and hence reversible transformation of the RF.

Since each application of the map $\cE_\alpha$ rotates the gyroscope towards the axis of polarization of the source particles $\langle \vec \pi \rangle$, the gyroscope will eventually line-up perfectly with the source particles. This happens in a time proportional to $\ell$.  In the Heisenberg picture, the cumulative effect of this reversible component of the  back-action after $t$ measurements is
\be
{e^\mu}_\alpha \xrightarrow{t} \uu^t_\alpha(e,\pi,1/\ell)({e^\mu}_\alpha).
\label{eq:deformt}
\ee
We note that rotations map coherent states to coherent states, so in leading order the RF remains in a coherent state.

The transformation $\cE_\alpha$ is not strictly unitary however, and non-unitary effects become manifest at higher orders. To understand this effect, it is helpful to think of the source particle as an ``environment" that couples to the RF. It is well understood that coupling to an environment typically implies a noisy non-unitary evolution of the RF \cite{Zur03a}. This is a consequence of the fact that after the measurement is performed, the source particle and the RF are in general entangled with one another.

Thus, the state of each gyroscopes will typically {\em deteriorate} due to the non-unitary effects of $\cE_\alpha$. As explained in Sec.~\ref{sec:approx}, the measurement of the quantum relational coordinate $\frak p_\alpha$ is in general an approximation of the semi-classical relational coordinate $p_\alpha$. The quality of the approximation depends on the state of the gyroscope --- more precisely it depends on the norm of the average polarization $\vec E_\alpha$. When a gyroscope is initially in a coherent state, this norm is maximal. As a consequence of the measurements of the source particles, $|\vec E_\alpha|$ will in general decrease, and thus the approximate equivalence between $p_\alpha$ and $\frak p_\alpha$ will deteriorate. This happens after a number of measurements proportional to $\ell$. At a later stage, when the unitary rotation has brought the RF in a state parallel to the source particles, $|\vec E_\alpha|$ will increase to become near maximal again. Figure 3 of Ref. \cite{PY07a} summarizes this dynamics.

Since the details of the dynamics are not important in the following analysis, we will assume that the essential features derived in Ref. \cite{PY07a} that we have just summarized extend beyond the case $j=\frac 12$. This hypothesis is corroborated by the fact that it is always possible to imagine a particle with spin $j>\frac 12$ as composed of $2j$ spin-$\frac 12$ particles in a symmetric state. Thus, the map induced on the RF by a spin-$j$ particle relates to the map obtained from $2j$ applications of the map induced by a spin-$\frac 12$ particle. Recent work \cite{BSLB07b} on the case $j>\frac 12$ also supports this picture.

\subsection{Possible experimental setting}
\label{sec:exp}

Now that we have described the dynamics incurred by the RF when used to consecutively measure the spin of source particles, we are in a position to determine whether this effect is detectable in principle. We do this by analyzing a {\em gendanken} experimental settings where signatures of such deformation could be detected.

The simplest experimental verification of the symmetry transformations of a RF uses spin-$\frac 12$ source particles and proceeds as follows:
\begin{enumerate}
\item Use the RF to measure the quantum relational coordinate of $t$ source particles.
\item Tabulate the statistics of the $\lambda^\pm_\alpha$ outcomes and compute $Pr_\alpha(\lambda^\pm)$.
\item Apply a rotation to the RF from its original position by an angle $\theta$, ${e^\mu}_\alpha \rightarrow {\Lambda^\alpha}_\beta {e^\mu}_\alpha$, and repeat step 1 and 2 for a different set of $t$ source particles, drawn from the same ensemble.
\end{enumerate}
According to Mallus' law, the probability of outcome of say $\lambda^+$ as a function of the angle $\theta$ should follow a cosine
\be
Pr_\alpha(\lambda^+,\theta) = \frac{1+Q\cos(\theta+\varphi_\alpha)}{2}
\label{eq:mallus}
\ee
where $\varphi_\alpha$ are some fixed  offsets and the visibility Q is between 0 and 1.

To verify this prediction within accuracy $\epsilon \ll 1$, one needs to perform each set of measurements on $t \approx 1/\epsilon^2$ source particles\footnote{The possibility of preparing the source particles in a massively entangled state could improve this bound to $t \approx 1/\epsilon$ \cite{GLM04a}. However, we assume here that the source particles are not controlled by the experimentalist,  for instance they could have an astrophysical origin. In that case, it can always be assumed that they are all in the same (possibly unknown) state $\sigma$ \cite{Ren07a}.}. These measurements will induce a back-action on the RF. If $t$ is too large, this back-action will completely deteriorate the RF, i.e. it will result in a visibility $Q=0$. Since this effect takes place after a number of measurements $\propto \ell$ \cite{PY07a}, it places a constraint on the size of the gyroscopes $\ell \gg 1/\epsilon^2$. In that regime, the dominant effect of the back-action will be to rotate the RF by an angle $\Delta\theta \approx t/\ell = 1/\ell\epsilon^2$, c.f. Eqs~(\ref{eq:U1}-\ref{eq:U2}).

Because of this additional rotation, the probability of the measurement outcomes in the subsequent setting will be altered. In the second round of measurements for instance, the measurement outcome probability is given by Eq.~\eqref{eq:mallus}, but with $\theta$ replaced by $\theta + \Delta\theta$ on the right hand side, where $\Delta\theta$ is the back-action rotation caused by the first set of measurements. This effect is only significant if the resulting probability differs from the predicted probability by an amount greater than the measurement accuracy $\epsilon$. To leading order, the correction to the probability is simply proportional to $\Delta\theta$, so we obtain the constraint $\Delta\theta \approx 1/\ell\epsilon^2 > \epsilon$.

Combining this with the constraint established in the previous paragraph, we conclude that the measurement back-action causes a rotation of the RF that is perceptible with accuracy $\epsilon \ll 1$ only if
\be
\frac{1}{\epsilon^2} \ll \ell < \frac{1}{\epsilon^3}.
\ee
This places important constraints on the regime where these effects are relevant, but demonstrate the possibility of detecting them in principle.

\section{Discussion}
\label{sec:discussion}

In this section, we discuss the connections and discrepancies between the directional RF model and deformed special relativity.

\subsection{Deformation}\label{sec:discussion1}
When a quantum directional RF is used to perform a constrained measurement on source particles, we have shown that
\begin{enumerate}
\item The RF undergoes a transformation $ {e^\mu}_\beta \rightarrow \cE_\alpha^\dagger({e^\mu}_\beta)$, where $\cE_\alpha$ is in general a non-linear function of the quantum relational coordinated $\frak p_\alpha$.
\item The map $\cE_\alpha$ has two components: a unitary $\uu_\alpha$ and hence reversible rotation, and a noisy {\em irreversible} component caused by the act of measurement that creates correlations between the RF and the source particles. The reversible effect is dominant when the RF is in a coherent state and its size $\ell \gg t$, where $t$ is the number of measurements.
\item The model we have analyzed admits a regime where these effects can be detected experimentally.
\end{enumerate}

The first point supports the idea that the quantum fluctuations incurred by RF can lead to an effective deformation of the symmetries of the relational observables. As opposed to the models investigated in \cite{LSV05a,AGG+05a,AGG+06a} however, this deformation is truly a dynamical effect. Indeed, to measure the relational coordinate $\frak p_\alpha$ requires somehow coupling the RF and the source particles. For instance, it can be achieved by turning on a Heisenberg interaction between the two systems and measuring their total energy, which would be directly proportional to the eigenvalues $\lambda_\alpha^k$ of $\frak p_\alpha$. Note moreover that it is really the coupling that causes the back-action, and the same dynamical map $\cE_\alpha$ would be obtained regardless of whether the total energy was actually measured. Thus, our conclusions are relatively insensitive to the details of the mixing between the RF and the source particles.

The second point is also crucial to the proposal of \cite{LSV05a,AGG+05a,AGG+06a}: only when the back-action has an inverse can it be interpreted as a symmetry deformation, c.f. Eqs.~(\ref{classicaltransf}-\ref{eq:nlt}), otherwise it is better described simply as noise. The question of whether a map is reversible or not on a subsystem has been studied in details in quantum information science, in the context of quantum error correction, see \cite{BNPV07a} for the latest developments. The question of {\em approximate} reversibility --- relevant to the present study --- has also been studied in the context of quantum error correction \cite{BK02a}.

While the leading effect of $\cE_\alpha$ on the RF was found to be reversible in our model, it also presented an important irreversible component. In fact, depending on the details of the experiment --- the number of times the RF is used, the polarization of the source particles, the size and initial state of the RF, etc --- this irreversible component can become the dominant effect of the back-action \cite{BRST06a}. In that case, the mixing between the RF and source particles would not yield a symmetry deformation; rather, it would reveal itself as a fundamental source of noise.

The existence of irreversible processes in an effective low-energy theory can be understood quite simply on general grounds. The effective theory is meant to describe the dynamics of the low-energy degrees of freedom once the high-energy degrees of freedoms have been traced out. In the presence of correlations between the low and high energy sectors --- which are to some extent inevitable in interacting theories --- the resulting dynamics will be non-unitary, with the high energy sector acting as an ``environment" decohering the low energy sector.  The approximate reversibility of the effective theory is a manifestation of the ``decoupling principle" according to which the dominant effect of the high energy theory on the low energy sector can be taken into account  by renormalizing its parameters.  Thus, according to this principle, the low energy sector is well approximated by a Hamiltonian, and hence reversible, dynamics.

The third point, regarding the measurability of the effect, is difficult to address in a more general context. This is partly due to the imperfect analogy between the model studied here and the situation faced by DSR. For instance, $SU(2)$ is compact while $SO(3,1)$ is not, DSR imposes a natural minimum mass $M_P$ while the small parameter $1/\ell$ in our model was tunable. These aspects will most likely alter the quantitative aspect of our analysis, and impact the analysis of Sec. \ref{sec:exp}.

Despite the common features shared between the gyroscope model and DSR, there exists an important caveat in the analogy. As mentioned above, the back-action on the RF is really a dynamical effect: the state of the RF after it has coupled to the source particle differs from its state prior to this interaction. The counterpart of Eq.~\eqref{umap} in the gyroscope model is thus
\be
{e^\mu}_\alpha(t) \xrightarrow{\uu_\beta} {e^\mu}_\alpha(t')
\ee
where $t$ is the time immediately before the measurement and $t'$ is the time immediately after, and $\uu_\alpha$ is the reversible approximation to $\cE_\alpha$. Contrarily to the proposal of \cite{LSV05a}, this does not directly imply a non-linear relation between the kinematical coordinate $\pi$ and its measured components $\frak p_\alpha$.

However, the value of $\frak p_\alpha$, like all quantum observables, can only be known probabilistically. Thus, as argued in Sec.~\ref{sec:exp}, a high precision measurement of the the relational coordinate requires repeating over a large ensemble of identically prepared source particles. Since each one of these measurement induces a non-linear back action on the RF, the ensemble average $\langle \frak p_\alpha \rangle_{ens}$ will be non-linearly related to the kinematical coordinate. Explicitly, this ensemble average is
\begin{eqnarray}
\langle \frak p_\alpha \rangle_{ens} &=& \frac 1t \sum_{i=t}^N \Tr\big\{\uu^{t-1}_\alpha({e^\mu}_\alpha) \pi_\mu \rho \otimes \sigma\big\} \\
&=& \Tr\big\{ \langle {E^\mu}_\alpha\rangle_{ens} \pi_\mu \sigma\big\}
\end{eqnarray}
where we have defined $\langle {E^\mu}_\alpha \rangle_{ens}$ as the time average of the kinematical coordinates of the RF $\langle {E^\mu}_\alpha \rangle_{ens} = \frac 1t \sum_t \Tr\{\uu^{t-1}_\alpha({e^\mu}_\alpha)  \rho\}$.  From an operational standpoint, we thus recover an analogue of Eq.~\eqref{umap}
\be
{e^\mu_\alpha} \xrightarrow{\langle \uu_\beta\rangle_{ens}} \langle {E^\mu}_\alpha\rangle_{ens}
\ee
only now the quantum average $\langle \cdot \rangle = \Tr\{ \cdot \rho\}$ used in the definition of ${E^\mu}_\alpha$, c.f. Eq.~\eqref{eq:Esu2}, needs to be supplemented by a dynamical average as induced by the back action. While this is an important conceptual difference, we note that the quantum average $\langle \cdot \rangle$ can only be estimated experimentally by repeating the measurement over a large ensemble, and so the dynamical average is inevitable.

\subsection{Multi-particle observables}\label{sec:discussion2}

The deformation of the Lorentz symmetry  implies a non-trivial construction of multi-particles observables. This is also the case in the gyroscope model, and the goal of this section is to briefly outline this analogy.

Different choices of deformation maps $\uu_{M_P}$ will lead to different momenta addition rules, which can be either non-commutative (and possibly also non-associative) or commutative (but still non-trivial). A typical example of the first class is the (associative) momenta addition arising in the bicrossproduct basis \cite{MR94a}:
\be\label{bicross}
p_0^{(tot)} = p_0^{(1)} + p_0^{(2)}, \quad p_i^{(tot)} =  e^{-p_0^{(2)}/M_P}p_i^{(1)} + p_i^{(2)}.\ee
Typically, non-commutative additions suffer from the ``soccer ball problem":  just like the single particle momentum $p_\alpha^{(i)}$, the total momentum of two particles $p_\alpha^{(tot)}$ is also bounded by the Planck mass. This is clearly in conflict with everyday observations where the total momentum of an object composed of a large number of particle, e.g. a soccer ball, can be much greater than $M_P$.

The commutative addition on the other hand can avoid this problem. A typical example of this class is obtained from the modified reference frame approach presented in Sec \ref{sec:DSR}. The intrinsic momentum of a 2-particles system is $\pi^{(tot)}=\pi^{(1)}+\pi^{(2)}$. In Minkowski space-time, the measured total momentum is simply
\be
\label{add}
p^{(tot)}_\alpha= \left(\pi^{(1)}+\pi^{(2)}\right)_\mu{e^\mu}_\alpha= p_\alpha^{(1)}+p_\alpha^{(2)}.
\ee
In the presence of a non-linear relation between the kinematical tetrad and its average, we obtain  instead
\be
\label{addeform}
p^{(tot)}_\alpha= \left(\pi^{(1)}+\pi^{(2)}\right)_\mu{E^\mu}_\alpha(e,(\pi^{(1)}+\pi^{(2)}), \gamma M_P),
\ee
where the rescaling factor $\gamma>1$ (related to the number of particles) is introduced  to avoid the soccer ball problem \cite{MS03a, GL04b}.

In the gyroscope model, the addition of vectors is straightforward: the total spin of two particles is
\be
\vec \pi^{(tot)}= \vec\pi^{(1)}+\vec\pi^{(2)}.
\ee
and so
\be
[\pi^{(1)}_\mu,\pi^{(2)}_\nu] =  0, \quad
[\pi^{(tot)}_\mu,\vec\pi^{(k)}_\mu]=0, \textrm{ with } k= 1,2.
\ee
The total relational coordinates is defined as before
\be
\mathfrak{p}_\alpha^{(tot)}= \pi^{(tot)}_\mu {e^{\mu}}_\alpha.
\ee
The quantum nature of the RF implies unusual features for these relational coordinates. For instance, the coordinates associated to distinct particles do not commute, e.g. for $\alpha = 1,2$,
\be
[\mathfrak{p}_\alpha ^{(1)},\mathfrak{p}_\alpha ^{(2)}] = \frac{{\epsilon_\rho}^{\mu\nu}}{\sqrt{\ell(\ell+1)}} \pi^{(1)}_\mu \pi^{(2)}_\nu {e_\alpha}^\rho
\ee
which vanishes when $\ell \rightarrow \infty$. This non-commutativity is in fact related to the measurement back-action: by the uncertainty principle, measuring the relational coordinate of the first particle will alter the value of the relational coordinate of the second particle. Similarly, the physical spin of each particles does not commute with their total spin $[\mathfrak{p}_\alpha^{(k)},\mathfrak{p}_\alpha^{(tot)}]\neq 0 $ for $k=1,2$. Again following the uncertainty principle,  measuring the total spin of two particles differs from measuring their individual spins and adding the outcomes.

While the gyroscope model is not plagued with the soccer ball problem, a similar problem arises when the RF is small compared to the size or number of measured particles. When $j > \ell$ for instance, the spectrum $\lambda^k$ of the quantum relational coordinate $\frak p_\alpha$, c.f. Eq.~\eqref{eq:eig}, is very different from the spectrum of the semi-classical coordinate $p_\alpha$. In fact, $p_\alpha$ always has $2j+1$ distinct eigenvalues while $\frak p_\alpha$ has at most $2\ell+1$ of them. Relatedly, the spacing between the eigenvalues of $\frak p_\alpha$ is $\lambda^{k+1} - \lambda^k \simeq 1 + O(\frac j\ell)$ as opposed to $1$ for the semi-classical coordinate, and this is noticeable when $j>\ell$.

Similarly, the total spin $\sum_i \frak p_\alpha^{(i)}$ of an arbitrary large collection of spin-$\frac 12$ particles relative to a fixed gyroscope can only take on a maximum number $n$ of distinct values. This contrasts with the kinematical spin $\pi_\mu$ or semi-classical relational spin $p_\alpha$ that both have a range proportional to the number of particles.  However, this ``soccer ball problem"  can be avoided by using a larger RF.
 
 %
%
%
%
%
%
%

\section{Conclusion}
\label{sec:conclusion}

In the light of recent analysis conducted in the context of quantum information science, we have examined the proposal that the back-action incurred by a quantum reference frame when used to measure a relational observable can lead to an effective deformation of its symmetries. This proposal was first put forth in a quantum gravitational setting \cite{LSV05a,AGG+05a,AGG+06a}, to support the idea that quantum fluctuations of the gravitational field can be described at an effective level by deformed special relativity. The existence of an auxiliary (kinematical) momentum variable carrying a linear realization of the Lorentz group and related non-linearly to the physical (relational) momentum is central to this proposal.

Our analysis led us to the conclusion that various forms of mixing between the RF and source particles --- and in particular mixing caused by the measurement of a relational observable --- will generally lead to a back-action on the RF that depends non-linearly on the relational coordinates. Since the back-action alters the state of the RF only {\em after} the measurement, it does not directly imply a non-linear relation between the kinematical and relational coordinates. However, a non-linear relation arises {\em on average}, when the RF is used to sequentially measure many source particles. Since repeated measurements are necessary to reach the accuracy required to probe these tiny effects, we obtain an effective non-linear relation between the kinematical and relational coordinates. This provides a concrete mechanism for the key component of the proposal \cite{LSV05a,AGG+05a,AGG+06a}.
If  this latter  is  taken seriously as  encoding DSR, our model shows that there is {\em a priori} no reason to restrict the symmetry deformation to the boost sector;  the rotations  could  also be  deformed. A DSR theory incorporating this feature should be considered and could provide new ways to experimentally constrain the deformation induced by quantum gravitational effects. We leave this question for future investigation.

Other analogies between our simple model and DSR were also briefly discussed. In particular,  multi-particles observables  share many similar non-trivial properties, like non-commutativity and spectrum saturation (a.k.a. the soccer ball problem in DSR). While these analogies are rather superficial, they may hint on a deeper connection worthy of further investigation.

Finally, the important and unavoidable role of irreversibility in the measurement back-action was also discussed. When used to perform sequential measurements on distinct particles, the RF will generally degrade. This places important constraints on the measurability of the effective symmetry deformation, which is caused by the reversible component of the back-action. In our simple model, there exist a regime in which the back-action is dominated by a its reversible component, but it is not clear how this generalizes to a broader context.

\section{Acknowledgment}

We thank Gerard Milburn, Kenny Pregnell, and Jon Yard for stimulating discussions that motivated parts of this work. DP is supported in part by the Gordon and Betty Moore Foundation through Caltech's Center for the Physics of Information, by the National Science Foundation under Grant No. PHY-0456720, and by the Natural Sciences and Engineering Research Council of Canada.  FG would like to thank M. Nielsen for his kind hospitality at the University of Queensland where this project was initiated.

\appendix
\section{Complete set of constrained observables}\label{appendixHO}

In this appendix, we construct the complete algebra of constrained observables for a collection of spin particles constrained to be invariant under a global $SU(2)$ symmetry. We recover in particular  the relational observables of Sec.~\ref{sec:RRF}.

This  algebra of constrained observables can be constructed using the Schwinger-Jordan representation which encodes the spin of a particle in a pair of constrained harmonic oscillators. Consider  two quantum harmonic oscillators:
$$
[a,a\dag]=[b,b\dag]=1, \quad [a,b]=0.
$$
These can be used to define spin operators
\begin{eqnarray}
&J_z=\frac{1}{2}(a\dag a-b\dag b), \qquad J_+=a\dag b,&\nonumber \\ & J_-=J_+\dag=ab\dag,\qquad E=\demi(a\dag
a+b\dag b).& \nonumber\end{eqnarray}
The operators $J$'s define a $su(2)$ algebra while the total energy $E$ commutes with the $J$'s and is thus a Casimir operator for
$su(2)$: \be [E,\cdot]=0, \qquad [J_z,J_\pm]=\pm J_\pm, \qquad [J_+,J_-]=2J_z. \nonumber\ee The spin representations at fixed $j$'s are given by fixing the
total energy $E$. Then diagonalising the two operators $E$ and $J_z$, we obtain the simple correspondence between the $SU(2)$ usual basis $|j m\rangle$
and the basis defined by the energy levels $|n_a n_b\rangle$ of the two oscillators:
\begin{eqnarray} &j=\demi(n_a+n_b),\quad m=\demi(n_a-n_b), & \nonumber\\
 &E\,|j m\rangle=j|j m\rangle, \quad J_z\,|j m\rangle=m|j m\rangle.& \nonumber\end{eqnarray}
Each particle $\pi_i$ can therefore be seen as arising from  a pair of harmonic oscillators $(a_i,b_i)$ with total energy fixed to $j$. Thus, we seek for the algebra of operators constructed from pairs of harmonic oscillators, with fixed energy, and invariant under the global $SU(2)$ action. It is given by \cite{GL05a} single-particle operators
$$
E_i=\demi\left(a_i^\dagger a_i+b_i^\dagger b_i\right),
$$
and operators acting on each pair of spins
\begin{eqnarray*}
E_{ij}&=&\demi\left(a_i^\dagger a_j+a_j^\dagger a_i+b_i^\dagger b_j+b_j^\dagger b_i\right), \\
F_{ij}&=&\frac{i}{2}\left(a_i^\dagger a_j-a_j^\dagger a_i+b_i^\dagger b_j-b_j^\dagger b_i\right).
\end{eqnarray*}
This way, we obtain $N+2N(N-1)/2=N^2$ operators. It is straightforward to check that the $E_i,E_{ij},F_{ij}$ are all Hermitian operators and form a
$u(N)$ Lie algebra. Then, quotienting by the trivially invariant operator $E=\sum_i E_i$, we get that the invariant algebra is ${su}(N)$. It is useful to introduce the following  operators $G_{ij}=E_{ij}+iF_{ij}$ and $G\dag_{ij}=E_{ij}-iF_{ij}=G_{ji}$, which have the  following commutation relations:
\begin{eqnarray} &[G_{ij},G\dag_{ij}]=2(E_j-E_i), \quad [G_{ij},E_i]=+\demi G_{ij},&\nonumber\\ &\quad [G_{ij},E_j]=-\frac{1}{2}G\dag_{ij}, \quad [G_{ij},G_{ki}]=G_{kj}
&\label{commutator}\nonumber \end{eqnarray}
The quantum relational coordinates for particle $k$ can be expressed in terms of these operators
\be
\mathfrak{p}_\alpha =\frac{1}{2}G^{\dag k\alpha}G^{k\alpha}-E^kE^\alpha-E^k.
\ee
As a final comment, note that this ``spin particles universe" can be interpreted as an intertwinner and associated to a fuzzy geometry \cite{GL05a}.


\end{document}